\documentclass[letterpaper, 10 pt, conference]{ieeeconf}  
\IEEEoverridecommandlockouts
\usepackage{graphicx} 

\usepackage{amsmath}
\usepackage{amsfonts}
\usepackage{amssymb}
\usepackage{amsthm}
\usepackage{xcolor}
\usepackage{caption}
\usepackage{subcaption}
\usepackage{cite}
\usepackage{url}
\theoremstyle{plain}

\newcommand{\frameworkname}{Safe Autonomy Run Time Assurance Framework}

\title{\LARGE \bf
A Universal Framework for Generalized Run Time Assurance with JAX Automatic Differentiation
}

\author{Umberto Ravaioli,$^{1}$ Kyle Dunlap,$^{2}$ and Kerianne Hobbs$^{3}$
\thanks{*Approved for Public Release, Case Numbers  AFRL-2022-3942. This work was supported by the Air Force Research Laboratory Innovation Pipeline Fund. The views expressed are those of the authors and do not reflect the official guidance or position of the United States Government, the Department of Defense or of the United States Air Force.}
\thanks{$^{1}$Umberto Ravaioli is with Toyon Research Corporation, Goleta, CA, 93117,
        {\tt\small uravaioli@toyon.com}}%
\thanks{$^{2}$Kyle Dunlap is with Parallax Advanced Research, Beavercreek, OH, 45431,
        {\tt\small kyle.dunlap@parallaxresearch.org}}%
\thanks{$^{3}$Kerianne Hobbs is on the Autonomy Capability Team (ACT3) at the Air Force Research Laboratory, Wright-Patterson AFB, OH, 45433,
        {\tt\small kerianne.hobbs@afrl.af.mil}}%
}

\begin{document}

\maketitle
\thispagestyle{empty}
\pagestyle{empty}

\begin{abstract}
With the rise of increasingly complex autonomous systems powered by black box AI models, there is a growing need for Run Time Assurance (RTA) systems that provide online safety filtering to untrusted primary controller output. Currently, research in RTA tends to be ad hoc and inflexible, diminishing collaboration and the pace of innovation. The \frameworkname{} presented in this paper provides a standardized interface for RTA modules and a set of universal implementations of constraint-based RTA capable of providing safety assurance given arbitrary dynamical systems and constraints. Built around JAX, this framework leverages automatic differentiation to populate advanced optimization based RTA methods minimizing user effort and error. To validate the feasibility of this framework, a simulation of a multi-agent spacecraft inspection problem is shown with safety constraints on position and velocity.
\end{abstract}


\section{INTRODUCTION}

As the mass deployment of autonomous systems begins to encroach upon safety critical domains, it will be necessary to develop robust and flexible safety assurance capabilities. While controllers were once designed by hand to be verifiably safe without intervention, this is not a feasible approach for the complex, black box algorithms of the future. Addressing this shortcoming, Run Time Assurance (RTA) \cite{schierman2020runtime} provides online safety filtering for untrusted primary controller, decoupling the safety problem from task completion. In doing so, RTA safety assurance techniques scale better as autonomous agents grow in complexity.

However, as RTA technology is in early stage development, the outputs of the research community have largely been ad hoc and hand designed. Software implementations are typically fully custom, lack portability to new scenarios, and require laborious derivative calculations. This holds back the growth of new RTA solutions and collaboration with other fields within autonomous systems. It is clear that the RTA community critically lacks a common toolset to standardize and streamline development.

In this paper, the Air Force Research Lab (AFRL) introduces the \frameworkname{} for building and deploying RTA systems in Python with minimal user effort. This framework provides a standard baseline for designing RTA modules, laying the foundation for interoperable and interchangeable RTA safety assurance within any organization's autonomy stack. The \frameworkname{} also provides constraint-based RTA modules with universal implementations of explicit/implict Simplex \cite{simplex_1} and explicit/implicit active set invariance filter (ASIF) \cite{gurriet2018online} RTA algorithms.

The constraint-based RTA modules are built around a set of generalized inequality constraint functions taking the form $h(\boldsymbol{x}) \ge 0$. By formulating an autonomous system's safety constraints in this manner, providing a system dynamics model, and, when relevant, a backup controller, the \frameworkname{} can produce a complete and operable RTA module for that system from any of the universal constraint-based RTA implementations. In brief, these modules will allow autonomous system designers to utilize RTA with minimal interaction with the underlying algorithms.

\begin{figure}[b!]
    \centering
    \includegraphics[width=1\columnwidth]{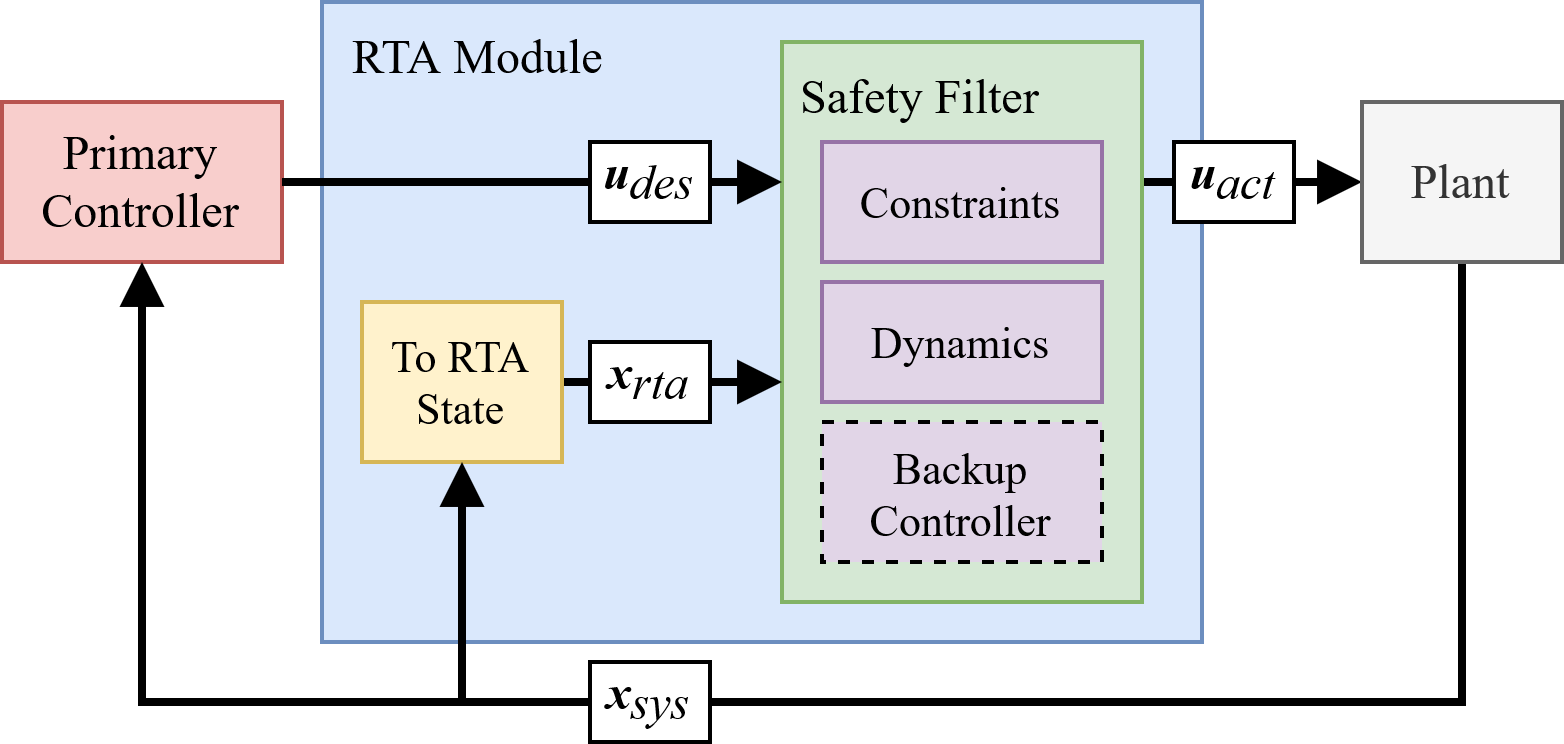}
    \caption{RTA module providing safety assurance to a feedback control system. The standardized RTA interface inputs/outputs are shown with white box labels on the data flow arrows. Note that the \textit{Constraints}, \textit{Dynamics}, and \textit{Backup Controller} boxes colored in purple are only required for universal constraint-based RTA implementations. Additionally, the \textit{Backup Controller} shown with a perforated border is only required for Simplex or implicit ASIF RTA modules.}
    \label{fig:RTA}
\end{figure}

To accomplish this, the constraint-based RTA modules utilize JAX \cite{jax2018github} to create differentiable constraints, dynamics, and backup controllers. These JAX implemented components can then be composed into the various Simplex and ASIF RTA implementations described in Section \ref{sec:rta}. By consistently utilizing differentiable JAX implementations, advanced ASIF techniques such as implicit ASIF and High-Order Control Barrier Functions (HOCBF) can be automatically supported and endlessly scaled with JAX automatic differentiation. This contribution in particular is key for proliferating these state-of-the-art and often opaque techniques to the wider autonomous system community.

The \frameworkname{} is released publicly on GitHub at \url{https://github.com/act3-ace/run-time-assurance} for anyone to use and contribute to. This repository includes a growing zoo of existing RTA implementations that utilize this framework to provide safety assurance to the autonomous control system models and tasks investigated by the AFRL Safe Autonomy Team.

\section{RUN TIME ASSURANCE}
\label{sec:rta}

The \frameworkname{} mainly focuses on control systems that are modeled as control affine dynamical systems, where a continuous-time system model is given by a system of ordinary differential equations,
\begin{equation} \label{eq:fxgu}
   \boldsymbol{\dot{x}} = f(\boldsymbol{x}) + g(\boldsymbol{x})\boldsymbol{u}.
\end{equation}
Here, $\boldsymbol{x} \in \mathcal{X} \subseteq \mathbb{R}^n$ denotes the state vector and $\boldsymbol{u}\in \mathcal{U} \subseteq\mathbb{R}^m$ denotes the control vector, where $\mathcal{X}$ defines the set of all possible state values and $\mathcal{U}$ defines the set of all admissible controls. For conciseness in this section, $\boldsymbol{x}$ is equivalent to $\boldsymbol{x}_{\rm rta}$.

RTA systems provide online safety filtering by separating the task of safety assurance from all other control objectives. As shown in Figure \ref{fig:RTA}, a feedback control system with RTA is split into a performance-focused \textit{primary controller} and a safety-focused \textit{RTA module}. In this figure, the primary controller is highlighted red to indicate low safety confidence, while the RTA module is highlighted blue to indicate high safety confidence. This structure allows the designer to isolate unverified components of the control system.

\subsection{Defining Safety}

For a dynamical system, safety at a given state $\boldsymbol{x}$ can be defined by a set of $M$ inequality constraints, $\varphi_i(\boldsymbol{x}): \mathcal{X} \to \mathbb{R}$,  $\forall i \in \{1,...,M\}$, where $\varphi_i(\boldsymbol{x}) \geq 0$ when the constraint is satisfied. The \textit{allowable set} $\mathcal{C}_{\rm A}$, is then defined as the set of states that satisfies all $M$ inequality constraints,
\begin{equation}
    \mathcal{C}_{\rm A} := \{\boldsymbol{x} \in \mathcal{X} \, | \, \varphi_i(\boldsymbol{x}) \geq 0, \forall i \in \{1,...,M\} \}.
\end{equation}
Note that $\mathcal{C}_{\rm A}$ only guarantees safety at the current state $\boldsymbol{x}(t_0)$, and not for states at future points in time. A state is said to be safe if it lies in a forward invariant subset of $\mathcal{C}_{\rm A}$ known as the \textit{safe set} $\mathcal{C}_{\rm S}$, where,
\begin{equation}
    \boldsymbol{x}(t_0) \in \mathcal{C}_{\rm S} \Longrightarrow \boldsymbol{x}(t) \in \mathcal{C}_{\rm A}, \forall t \geq t_0.
\end{equation}

In dynamical control systems, the control input $\boldsymbol{u}$ is bounded by the admissible control set $\mathcal{U}$. Therefore $\mathcal{C}_{\rm S}$ must also be a control invariant subset of $\mathcal{C}_{\rm A}$, where there exists a control law $\boldsymbol{u} \in \mathcal{U}$ that renders $\mathcal{C}_{\rm S}$ forward invariant. For control systems, $\mathcal{C}_{\rm S}$ can be defined both \textit{explicitly} and \textit{implicitly}, where safety can be assured for all time. 

First, $\mathcal{C}_{\rm S}$ can be defined explicitly by a set of $M$ control invariant inequality safety constraints, $h_i(\boldsymbol{x}): \mathcal{X}\to \mathbb{R}$,  $\forall i \in \{1,...,M\}$, where again $h_i(\boldsymbol{x}) \geq 0$ when the constraint is satisfied. $\mathcal{C}_{\rm S}$ is then defined as,
\begin{equation} \label{eq: explicitly_defined_safe_set}
    \mathcal{C}_{\rm S} := \{\boldsymbol{x}\in\mathcal{X} \, | \, h_i(\boldsymbol{x})\geq 0, \forall i \in \{1,...,M\} \}.
\end{equation} 

$\mathcal{C}_{\rm S}$ can also be defined implicitly using closed loop trajectories under a backup control law $\boldsymbol{u}_{\rm b}$, where $\boldsymbol{u}_{\rm b}$ directs the system to a verified backup set $\mathcal{C}_{\rm B} \subseteq \mathcal{C}_{\rm S}$, where the system is known to remain safe for all time. $\mathcal{C}_{\rm S}$ is then defined as,
\begin{equation}  \label{eq: implicitly_defined_safe_set}
    \mathcal{C}_{\rm S} := \{\boldsymbol{x}\in\mathcal{X} \, | \, \forall t\geq 0,\,\, \phi^{\boldsymbol{u}_{\rm b}}(t;\boldsymbol{x})\in\mathcal{C}_{\rm A} \},
\end{equation}
where $\phi^{\boldsymbol{u}_{\rm b}} $ is defined as a prediction of the state $\boldsymbol{x}$ for $t$ seconds under $\boldsymbol{u}_{\rm b}$. Note that $\mathcal{C}_{\rm S}$ can be calculated entirely offline when defined explicitly, but it must be calculated online at each state when defined implicitly.

\subsection{Simplex Algorithms}

One type of RTA filter is the Simplex filter, which switches between primary and backup controllers to assure safety of the system. The Simplex filter monitors the desired control input $\boldsymbol{u}_{\rm des}$ from the primary controller, predicts the next state of the system, and evaluates if this state is safe or not. If it is safe, $\boldsymbol{u}_{\rm des}$ is passed to the plant unaltered as $\boldsymbol{u}_{\rm act}$. Otherwise, a backup control input $\boldsymbol{u}_{\rm b}$ is passed to the plant. For this framework, the Simplex filter is constructed as follows.
\noindent \rule{1\columnwidth}{0.7pt}
\noindent \textbf{Simplex Filter}
\begin{equation}
\begin{array}{rl}
\boldsymbol{u}_{\rm act}(\boldsymbol{x})=
\begin{cases} 
\boldsymbol{u}_{\rm des}(\boldsymbol{x}) & {\rm if}\quad \phi_1^{\boldsymbol{u}_{\rm des}}(\boldsymbol{x}) \in \mathcal{C}_{\rm S}  \\ 
\boldsymbol{u}_{\rm b}(\boldsymbol{x})  & {\rm if}\quad otherwise
\end{cases}
\end{array}\label{eq:switching}
\end{equation}
\noindent \rule[7pt]{1\columnwidth}{0.7pt}
Here, $\phi_1^{\boldsymbol{u}_{\rm des}}(\boldsymbol{x})$ represents a prediction of the state $\boldsymbol{x}$ when $\boldsymbol{u}_{\rm des}$ is applied for one discrete time interval, which is found using the system dynamics in Eq. \eqref{eq:fxgu}.
Simplex filters can be defined either explicitly or implicitly, where $\mathcal{C}_{\rm S}$ is defined using Eq. \eqref{eq: explicitly_defined_safe_set} or \eqref{eq: implicitly_defined_safe_set} respectively.

\subsection{ASIF Algorithms}

Another type of RTA filter is ASIF, which is an optimization-based technique designed to minimize deviation from the primary controller while still assuring safety. ASIF algorithms are based on the use of control barrier functions \cite{ames2019control} 
to enforce safety, and use a quadratic program to minimize the $l^2$ norm difference between $u_{\rm des}$ and $u_{\rm act}$. For this framework, the ASIF algorithm is constructed as follows.
\noindent \rule{1\columnwidth}{0.7pt}
\noindent \textbf{Active Set Invariance Filter}
\begin{equation}
\begin{gathered}
\boldsymbol{u}_{\text{act}}(\boldsymbol{x})={\text{argmin}}  \left\Vert \boldsymbol{u}_{\text{des}}-\boldsymbol{u}\right\Vert ^{2}\\
\text{s.t.} \quad  \boldsymbol{u} \in \mathcal{U}, \quad BC_i(\boldsymbol{x},\boldsymbol{u})\geq 0, \quad \forall i \in \{1,...,M\}
\end{gathered}\label{eq:optimization}
\end{equation}
\noindent \rule[7pt]{1\columnwidth}{0.7pt}
Here, $BC_i(\boldsymbol{x},\boldsymbol{u})$ represents a set of $M$ barrier constraints designed to assure safety. These barrier constraints enforce Nagumo's condition \cite{nagumo1942lage}, where the boundary of the set formed by $h_i(\boldsymbol{x})$ is examined to ensure $\dot{h}_i(\boldsymbol{x}) \geq 0$, causing $\boldsymbol{x}$ to never leave $\mathcal{C}_{\rm S}$. For the $i^{th}$ constraint, this condition is written as,
\begin{equation}
    \dot{h}_i(\boldsymbol{x}) = \nabla h_i(\boldsymbol{x}) \dot{\boldsymbol{x}} = L_f h_i(\boldsymbol{x}) + L_g h_i (\boldsymbol{x}) \boldsymbol{u} \geq 0,
\end{equation}
where $L_f$ and $L_g$ are Lie derivatives of $h_i$ along $f$ and $g$ respectively. However, it is not practical to enforce this condition on its own, as the boundary of $\mathcal{C}_{\rm S}$ has no volume. Therefore, the constraint is modified to include a class $\kappa$ strengthening function $\alpha(x)$, which is continuous, strictly increasing, has the condition $\alpha(0)=0$, and is used to relax the barrier constraint away from the boundary of $\mathcal{C}_{\rm S}$. The barrier constraint is then defined as,
\begin{equation}
    BC_i(\boldsymbol{x},\boldsymbol{u}) := L_f h_i(\boldsymbol{x}) + L_g h_i (\boldsymbol{x}) \boldsymbol{u} + \alpha(h_i(\boldsymbol{x})) \geq 0.
\end{equation}

ASIF RTAs, like Simplex RTAs, can also be defined explicitly or implicitly. Using the system dynamics in Eq. \eqref{eq:fxgu}, each barrier constraint can first be defined explicitly as,
\begin{equation} \label{eq:exp_BC}
    BC_i(\boldsymbol{x},\boldsymbol{u}) := \nabla h_i(\boldsymbol{x}) (f(\boldsymbol{x}) + g(\boldsymbol{x})\boldsymbol{u}) + \alpha(h_i(\boldsymbol{x})),
\end{equation}
where, $h_{i}(\boldsymbol{x})$ refers to the set of control invariant inequality safety constraints. For some systems, $\nabla h(\boldsymbol{x})$ may not depend on $\boldsymbol{u}$, and therefore Eq. \eqref{eq:exp_BC} is no longer a valid barrier constraint because the term $\nabla h(\boldsymbol{x})  g(\boldsymbol{x})\boldsymbol{u}$ vanishes from the equation. Therefore, a sequence of inequality constraints $\Psi_i(\boldsymbol{x}), \forall i \in \{1,...,m\}$ must be defined such that $\dot{\Psi}_m(\boldsymbol{x})g(\boldsymbol{x}) \neq 0$, where $m$ is referred to as the relative degree of the system \cite{xiao2022control}. These constraints, known as HOCBFs, are defined as,
\begin{equation} \label{eq:HOCBF}
    \Psi_i(\boldsymbol{x}) := \dot{\Psi}_{i-1}(\boldsymbol{x}) + \alpha_i(\Psi_{i-1}(\boldsymbol{x})), \quad \forall i \in \{1,...,m\},
\end{equation}
where $\Psi_0(\boldsymbol{x})=h(\boldsymbol{x})$, and in the cases where $i < m$, $\dot{\Psi}(\boldsymbol{x})=\nabla \Psi(\boldsymbol{x}) f(\boldsymbol{x})$. Note that when the relative degree of the system is 1 ($m=1$),  $\Psi_1(\boldsymbol{x})$ is equivalent to the barrier constraint developed in Eq. \eqref{eq:exp_BC}.

Each barrier constraint can also be defined implicitly as,
\begin{equation}\label{eq:imp_BC}
    BC_i(\boldsymbol{x},\boldsymbol{u}) := \nabla \varphi_i(\phi^{\boldsymbol{u}_b}_j) D(\phi^{\boldsymbol{u}_b}_j) (f(\boldsymbol{x}) + g(\boldsymbol{x})\boldsymbol{u}) + \alpha(\varphi_i(\phi^{\boldsymbol{u}_b}_j) )
\end{equation}
where $\varphi_i(\boldsymbol{x})$ refers to the set of inequality constraints that define $\mathcal{C}_{\rm A}$, $\phi^{\boldsymbol{u}_b}_j$ refers to the state at the $j^{th}$ discrete time interval along the backup trajectory $\forall t \in [0,T]$, and $D(\phi^{\boldsymbol{u}_b}_j)$ is computed by integrating a sensitivity matrix differential equation along the backup trajectory \cite{gurriet2020scalable}.

\section{UNIVERSAL FRAMEWORK}
The \frameworkname{} is a modular toolset for building RTA capability into arbitrary autonomous/control systems. The following section describes the design of this framework and the details of its universal architecture.

\subsection{Interface}
The \frameworkname{} standardizes the interface for RTA modules as shown in Figure \ref{fig:RTA}. RTA modules use a desired input action $\boldsymbol{u}_{\rm des}$ and the current system state $\boldsymbol{x}_{\rm sys}$ to produce a safe output action $\boldsymbol{u}_{\rm act}$. Any arbitrary code/algorithm that adheres to these inputs and outputs can be wrapped in the base RTA interface to create a \frameworkname{} compatible RTA module.

The control input and outputs are 1D NumPy arrays corresponding to standard control vectors. However, the system state can be any arbitrary python object that encapsulates the current state. Within the RTA module, the arbitrary system state is converted to an RTA state, $\boldsymbol{x}_{\rm rta}$, that is composed of a 1D NumPy array. By default, the system state is assumed to be a 1D NumPy array and is directly passed as the RTA state, however any arbitrary custom conversion code can be used instead. The converted state vector and the input control vector are then passed to the internal safety filter.

This interface flexibility allows creation of arbitrary safety filters as RTA modules that can be seamlessly slotted in or composed. An additional RTA module enabled by this capability is the included \textit{cascaded RTA} module which is a wrapper for a sequential cascade of internal RTA modules that apply RTA filtering of increasing priority to the output of the previous module in the sequence. Following this design pattern, additional RTA composition patterns can be iterated endlessly.

\subsection{Constraint-Based RTA}
The \frameworkname{} includes a number of universal Simplex and ASIF constraint-based RTA implementations that are based around a common set of constraint, dynamics, and backup controller interfaces. By implementing custom versions of these interfaces for a given problem, one enables turn-key instances of these RTA algorithms. All the constraint-based RTA implementations are based on JAX, which is described below in Section \ref{sec:jax}, to allow for composable automatic differentiation of RTA algorithm components.
\subsubsection{Constraints} The \frameworkname{} includes constraint classes for defining safety constraints in terms of a monotonic function $h(\boldsymbol{x}_{\rm rta})$ of the RTA State (Note: $h$ is interpreted as $\varphi$ for implicit methods). Safety is defined when $h(\boldsymbol{x}_{\rm rta}) \ge 0$. Constraints may also be associated with constraint strengthening functions $\alpha(h(\boldsymbol{x}_{\rm rta}))$. Through evaluation and manipulation of associated constraints, constraint-based RTA modules maintain system safety.
\subsubsection{Dynamics} Constraint-based RTA requires knowledge of the system dynamics. This allows it to determine safe actions that will drive the system away from constraint violation boundaries. Note that the dynamics model need not exactly estimate the true system dynamics but needs to model the potentially simplified RTA state. There are two types of dynamics information that may be needed by constraint-based RTA: next state propagation and state transition derivatives. Note that implicit ASIF requires both types of dynamics to be implemented.
\begin{itemize}
    \item \textit{Next State Propagation}: This dynamics form is used by Simplex and implicit ASIF to estimate the next RTA state given a current RTA state, control vector, and time interval. Note that for implicit methods, a high fidelity dynamics model is recommended for long-term trajectory estimation. For this reason, the next state propagation is not required to be implemented in JAX and is not differentiated allowing for high fidelity external simulators to be used.
    \item \textit{State Transition Derivative}: ASIF methods require a knowledge of the instantaneous state time derivative of the form $\dot{\boldsymbol{x}}_{\rm rta} = f(\boldsymbol{x}_{\rm rta}) + g(\boldsymbol{x}_{\rm rta})u$. The $f(\boldsymbol{x}_{\rm rta})$ and $g(\boldsymbol{x}_{\rm rta})u$ components must be implemented separately for use in different parts of the ASIF algorithms.
\end{itemize}
\subsubsection{Backup Controller} Simplex and implicit ASIF RTA algorithms rely on backup controllers that can drive the system to a known safe backup set. The \frameworkname{} provides a backup controller class for implementing differentiable backup controllers. Internal states of the backup controller (e.g. integral error or moving set points) are supported but must be maintained within a NumPy array or dictionary of NumPy arrays in order to maintain JAX differentiability and compilation. The internal state can be saved and restored to enable backup trajectory computation in implicit RTA methods. Explicit ASIF RTA does not require a backup controller.

\subsection{JAX}
\label{sec:jax}
JAX \cite{jax2018github} is an experimental Python library from Google for performing automatic differentiation via Autograd and Accelerated Linear Algebra (XLA) Just-In-Time (JIT) compilation of native NumPy and Python code. Autograd, the technology that underlies most Deep Learning frameworks, enables efficient point value derivative computation of arbitrary functions. JIT compilation results in massive execution time benefits, especially relevant for complicated derivatives, by automatically converting native Python code into optimized compiled programs at runtime. Note that Autograd is \textit{not} the same as symbolic differentiation or numerical differentiation, by only computing single point values it is highly performant, exact, and runs in constant time.

JAX lies at the heart of The \frameworkname{}'s universal nature, particularly for the powerful ASIF methods. The included constraint-based RTA implements constraints, dynamics state transition derivatives, backup control, and algorithm specific logic entirely in JAX. Instead of requiring derivatives to be manually implemented, they are automatically computed, eliminating additional user effort and error. This is especially useful for HOCBFs and implicit ASIF methods that require large amounts of differentiation that may include long product-rule chains. Without JAX Autograd, implementing these techniques becomes tedious and error-prone.

\section{SIMULATION}

This section develops a simulation of a dynamical control system to demonstrate the \frameworkname{}'s feasibility and usability. The simulation considers a multi-agent spacecraft inspection problem, where multiple active ``deputy" spacecraft examine a passive ``chief" spacecraft. For this simulation, explicit ASIF RTA is used to assure safety, where each deputy spacecraft uses a separate instance of RTA. While creating an explicit ASIF RTA filter may traditionally be a difficult task, with this framework the user needs to define the admissible control set $\mathcal{U}$, the functions $f(\boldsymbol{x})$ and $g(\boldsymbol{x})$ from Eq. \eqref{eq:fxgu}, the control invariant safety constraints $h_i(\boldsymbol{x})$, and a strengthening function $\alpha(x)$ for each constraint.

\subsection{Spacecraft Dynamics}

\begin{figure}[bt!]
\centering
\includegraphics[width=.35\textwidth]{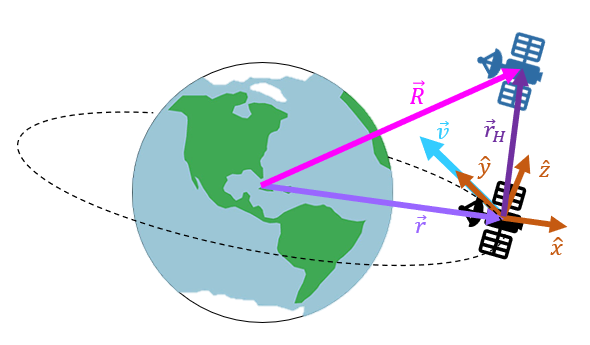}
\caption{Hill's reference frame centered on a chief spacecraft.}
\label{fig:HillsFrame}
\end{figure}

For the spacecraft inspection problem, Hill's reference frame \cite{hill1878researches} is used to represent the location of each spacecraft. As shown in Figure \ref{fig:HillsFrame}, the origin of Hill's frame is located at the center of mass of the chief, the unit vectors $\hat{x}$ and $\hat{y}$ point away from the center of the Earth and in the direction of motion of the chief respectively, and the unit vector $\hat{z}$ is normal to $\hat{x}$ and $\hat{y}$. The linearized relative motion dynamics between the $i^{th}$ deputy and the chief are given by the Clohessy-Wiltshire equations \cite{clohessy1960terminal}, 
\begin{equation} \label{eq: system dynamics}
    \dot{\boldsymbol{x}}_i = A {\boldsymbol{x}_i} + B\boldsymbol{u},
\end{equation}
where the state $\boldsymbol{x}_i=[x,y,z,\dot{x},\dot{y},\dot{z}]^T \in \mathcal{X}=\mathbb{R}^6$, the control $\boldsymbol{u}= [F_x,F_y,F_z]^T \in \mathcal{U} = [-u_{\rm max},u_{\rm max}]^3$, and
\begin{align}
\centering
    A = 
\begin{bmatrix} 
0 & 0 & 0 & 1 & 0 & 0 \\
0 & 0 & 0 & 0 & 1 & 0 \\
0 & 0 & 0 & 0 & 0 & 1 \\
3n^2 & 0 & 0 & 0 & 2n & 0 \\
0 & 0 & 0 & -2n & 0 & 0 \\
0 & 0 & -n^2 & 0 & 0 & 0 \\
\end{bmatrix}, 
    B = 
\begin{bmatrix} 
 0 & 0 & 0 \\
 0 & 0 & 0 \\
 0 & 0 & 0 \\
\frac{1}{m} & 0 & 0 \\
0 & \frac{1}{m} & 0 \\
0 & 0 & \frac{1}{m} \\
\end{bmatrix}.
\end{align}
Here, $m$ is the mass of the deputy and $n$ is the mean motion of the chief's orbit. For $N$ deputies, it is assumed that the full system state consists of the state of each deputy, where only one deputy is controlled at a time. Therefore $\boldsymbol{x}$ becomes a vector of length $6*N$, where $\boldsymbol{x}=[\boldsymbol{x}_1, ... , \boldsymbol{x}_N]$. 
The functions $f(\boldsymbol{x})$ and $g(\boldsymbol{x})$ are then size $(6*N) \times (6*N)$ and $(6*N) \times 3$ respectively, and are defined as,
\begin{equation}
    f(\boldsymbol{x})=
    \begin{bmatrix}
     A & \boldsymbol{0}_{6\times 6}  & \dots \\
     \boldsymbol{0}_{6\times 6} & A &  \\
     \vdots & & \ddots  \\
    \end{bmatrix}, \quad
    g(\boldsymbol{x})=
    \begin{bmatrix}
     B \\
     \boldsymbol{0}_{6\times 3} \\
     \vdots \\
    \end{bmatrix}.
\end{equation}

\subsection{Safety Constraints} \label{sec:safety_constraints}

The spacecraft inspection problem considers several safety constraints based on position and velocity
\cite{hobbs2020elicitation}
\cite{hobbs2021risk}.
Note that in order for the constraints to be used with explicit ASIF RTA, they need to be control invariant. All proofs for control invariance can be found in the papers \cite{dunlap2021comparing} and \cite{hibbard2022guaranteeing}. The following constraints are enforced for $N$ deputies, $\forall i \in \{1,...,N\}$.

\subsubsection{Chief Collision Avoidance}

Each deputy shall not collide with the chief. This constraint is defined as,
\begin{equation}
    \varphi_1(\boldsymbol{x}) := \Vert \boldsymbol{p}_i \Vert_2 - (r_{\rm d}+r_{\rm c}) \geq 0,
\end{equation}
where $\boldsymbol{p}=[x, y, z]^T$, $r_{\rm d}$ is the collision radius of each deputy, and $r_{\rm c}$ is the collision radius of the chief.

\subsubsection{Deputy Collision Avoidance}

Each deputy shall not collide with any other deputies. $\forall \, j \in \mathbb{Z}_{1:N}$, $i\neq j$, this constraint is defined as,
\begin{equation}
    \varphi_2(\boldsymbol{x}) := \Vert \boldsymbol{p}_i - \boldsymbol{p}_j \Vert_2 - 2r_{\rm d} \geq 0.
\end{equation}

\subsubsection{Dynamic Speed Constraint}

The speed of each deputy shall decrease as it moves closer to the chief. This reduces risk of a high speed collision, as well as risk in the event of a fault \cite{mote2021natural}. This constraint is defined as,
\begin{equation}
    \varphi_3(\boldsymbol{x}) := \nu_0 + \nu_1\Vert \boldsymbol{p}_i \Vert_2 - \Vert \boldsymbol{v}_i \Vert_2 \geq 0,
\end{equation}
where $\nu_0$ is a minimum allowable docking speed, $\nu_1$ is a constant rate at which $\boldsymbol{p}$ shall decrease, and $\boldsymbol{v}=[\dot{x}, \dot{y}, \dot{z}]^T$.

\subsubsection{Sun Avoidance}

Assuming each deputy is pointing sensors directly at the chief, these sensors shall not align with the sun. This constraint is defined as,
\begin{equation}
    \varphi_4(\boldsymbol{x}) := -\frac{\langle \boldsymbol{p}_i, \hat{e}_s \rangle}{\Vert \boldsymbol{p}_i \Vert_2} + \cos \frac{\theta_s}{2} \geq 0,
\end{equation}
where $\hat{e}_s$ is a unit vector pointing from the sun to the chief and $\theta_s$ is an angle defining a conic exclusion zone that the deputy shall not align with.

\subsubsection{Aggressive Maneuvering}

Each deputy shall not maneuver aggressively with high velocities. This is defined in terms of three separate constraints,
\begin{equation}
\begin{gathered}
    \varphi_5(\boldsymbol{x}) := v_{\rm max}^2 - \dot{x}_i^2\geq 0, \quad \varphi_6(\boldsymbol{x}) := v_{\rm max}^2 - \dot{y}_i^2\geq 0, \\ \varphi_7(\boldsymbol{x}) := v_{\rm max}^2 - \dot{z}_i^2\geq 0,
\end{gathered}
\end{equation}
where $v_{\rm max}$ is the maximum allowable velocity.

\subsubsection{Actuation Saturation}

Each deputy shall remain within the bounds of its actuation limits. This is not implemented as an RTA constraint, but rather as an inequality constraint to the quadratic program, such that $\boldsymbol{u} \in \mathcal{U}$.

Note that for all constraints, a strengthening function $\alpha(x)$ must be provided. For this simulation, $\alpha(x)=10^ax+10^bx^3$, where $a, b \in [-3, -1]$.

\subsection{Results}

\begin{figure*}[tb!]
    \begin{subfigure}[t]{0.49\columnwidth}
        \centering
        \includegraphics[width=\linewidth]{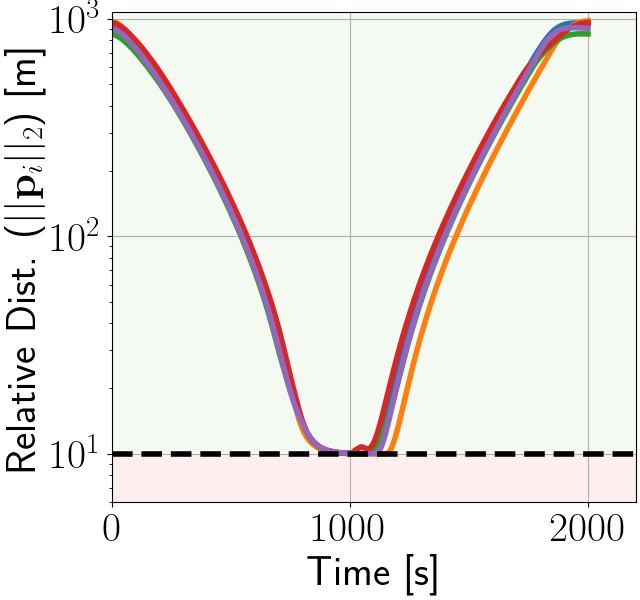}
        \caption{Chief collision avoidance.}
        \label{fig:chief_collision}
    \end{subfigure}
    \begin{subfigure}[t]{0.49\columnwidth}
        \centering
        \includegraphics[width=\linewidth]{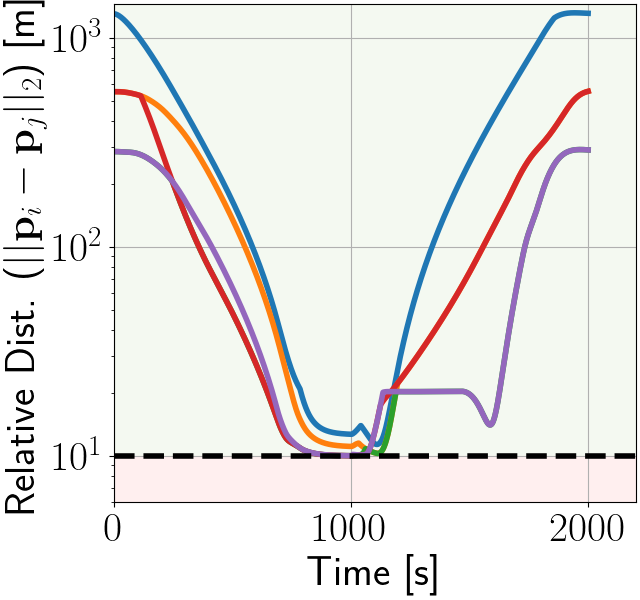}
        \caption{Deputy collision avoidance.}
        \label{deputy_collision}
    \end{subfigure}
    \begin{subfigure}[t]{0.49\columnwidth}
        \centering
        \includegraphics[width=\linewidth]{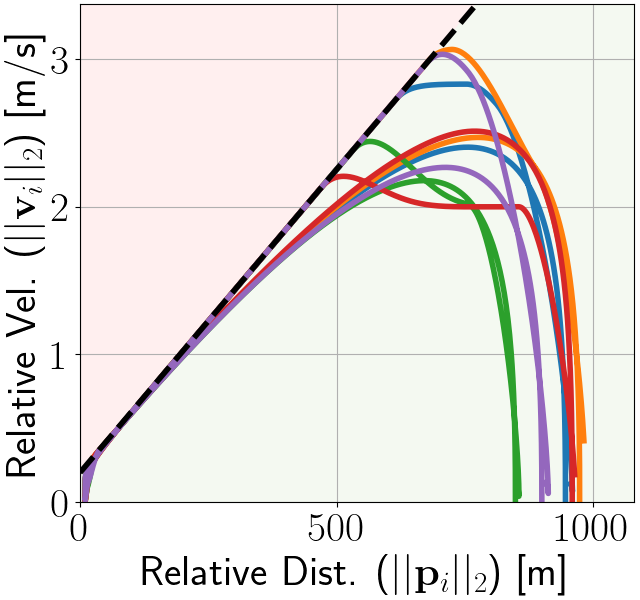}
        \caption{Dynamic speed constraint.}
        \label{fig:dynamic_speed}
    \end{subfigure}
    \begin{subfigure}[t]{0.49\columnwidth}
        \centering
        \includegraphics[width=\linewidth]{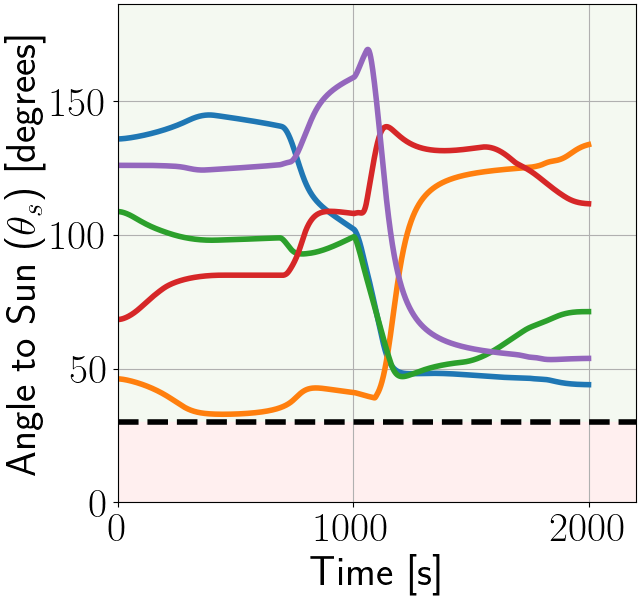}
        \caption{Sun avoidance.}
        \label{fig:sun_avoid}
    \end{subfigure}
    \begin{subfigure}[t]{0.49\columnwidth}
        \centering
        \includegraphics[width=\linewidth]{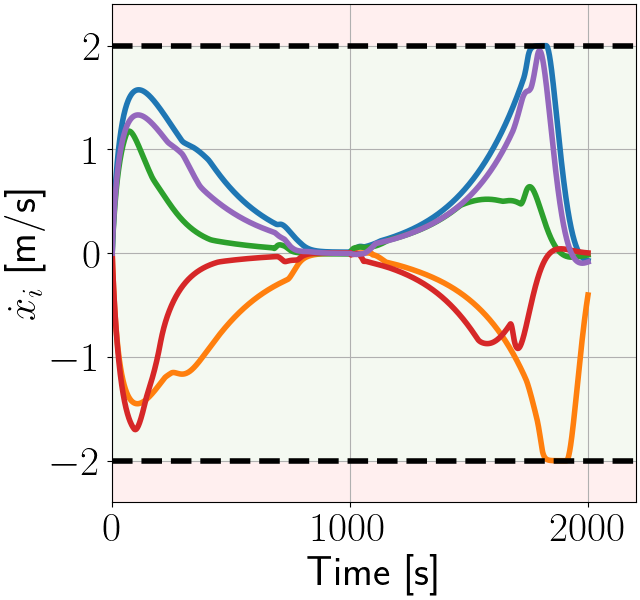}
        \caption{Aggressive maneuvering: $\dot{x}$.}
        \label{fig:xdot_lim}
    \end{subfigure}
    \begin{subfigure}[t]{0.49\columnwidth}
        \centering
        \includegraphics[width=\linewidth]{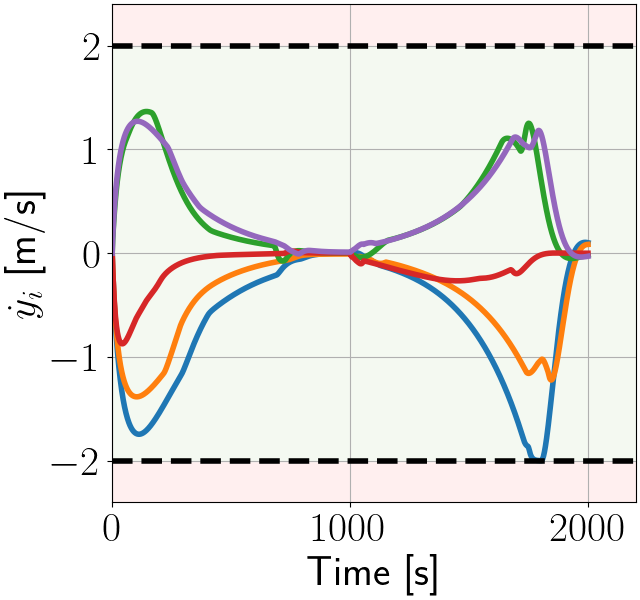}
        \caption{Aggressive maneuvering: $\dot{y}$.}
        \label{ydot_lim}
    \end{subfigure}
    \begin{subfigure}[t]{0.49\columnwidth}
        \centering
        \includegraphics[width=\linewidth]{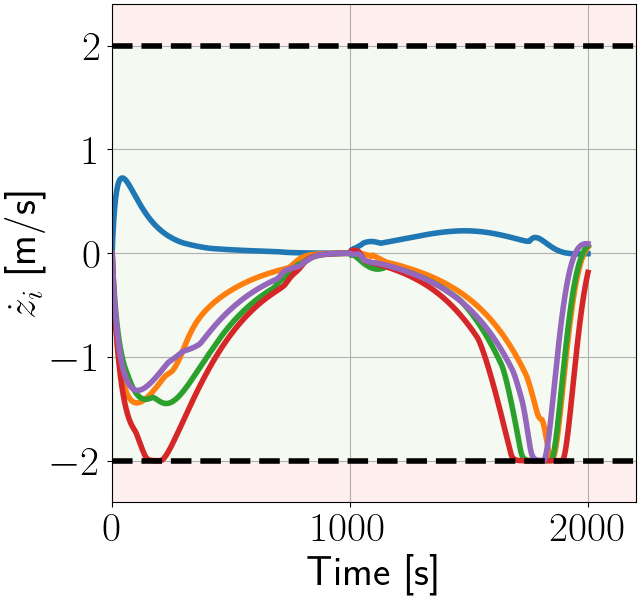}
        \caption{Aggressive maneuvering: $\dot{z}$.}
        \label{fig:zdot_lim}
    \end{subfigure}
    \begin{subfigure}[t]{0.49\columnwidth}
        \centering
        \includegraphics[width=\linewidth]{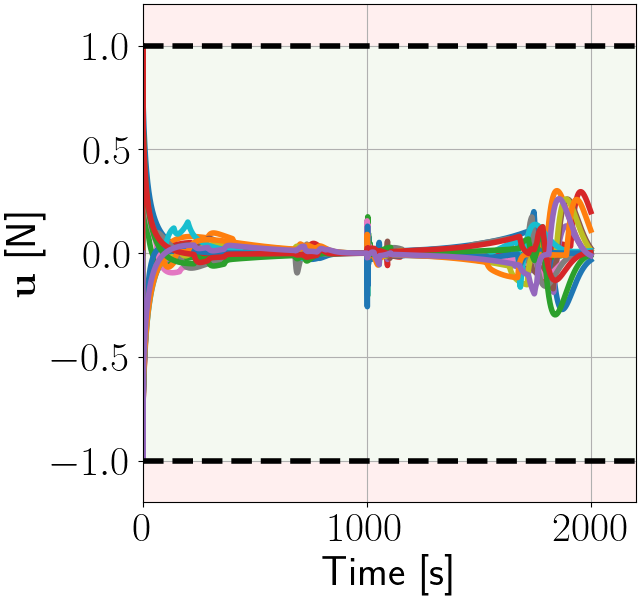}
        \caption{Actuation saturation.}
        \label{fig:actuation}
    \end{subfigure}
    \caption{Explicit ASIF simulation results.}
    \label{fig:sim}
    \centering
\end{figure*}

Using the dynamics and constraints developed in the previous sections, an explicit ASIF RTA filter can be created. To simulate the system, LQR is used as the primary controller, where it is designed to be aggressive and violate the safety constraints to show that RTA effectively assures safety. The simulation is run for 2,000 seconds, where the time intervals are 1 second. Five deputies are used in the simulation, where each is controlled by a separate but identical LQR controller and RTA filter. The simulation results are shown in Figure \ref{fig:sim}, where all constraints from Section \ref{sec:safety_constraints} are shown. In each sub-figure, the regions shaded red represent $\varphi_i(\boldsymbol{x}) < 0$, the regions shaded green represent $\varphi_i(\boldsymbol{x}) > 0$, and the black dashed lines represent $\varphi_i(\boldsymbol{x}) = 0$. Each deputy is represented by a solid, colored line.
The parameters used for this simulation are: $u_{\rm max}=1$ N, $m=12$ kg, $n=0.001027$ rad/s, $r_{\rm d}=5$ m, $r_{\rm c}=5$ m, $\nu_0=0.2$ m/s, $\nu_1=4*n$ 1/s, $\hat{e}_s=[1, 0, 0]$, $\theta_s=\pi/6$ rad, and $v_{\rm max}=2$ m/s.



Figure \ref{fig:sim} shows that the explicit ASIF RTA simultaneously assures safety of all constraints for the entire simulation. In total, there are 5 instances of RTA, where each instance is enforcing 10 constraints ($\varphi_2(\boldsymbol{x})$ is counted 4 times). This results in 50 gradient computations per time interval, many of which are non-trivial to compute analytically. The simulation was run with an 11th Gen 3.00 GHz Intel Core i7-1185G7 CPU with 16 GB RAM, where the average computation time across 100 simulations was 1.43 seconds per simulation.

\subsection{HOCBF Simulation}

\begin{figure}[htb!]
    \centering
    \includegraphics[width=.55\columnwidth]{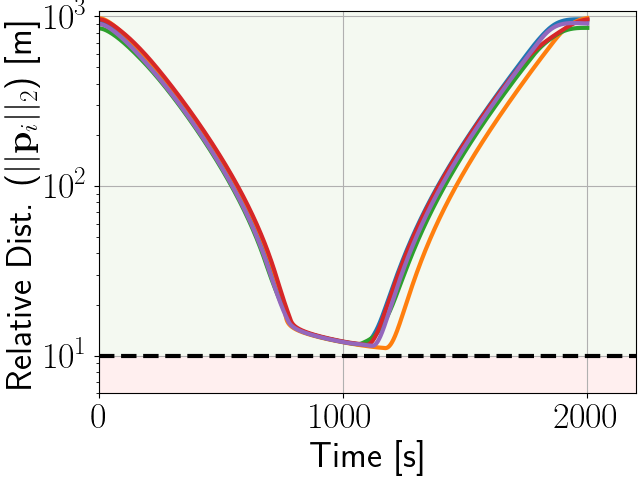}
    \caption{Chief collision avoidance constraint, using a HOCBF.}
    \label{fig:hocbf_collision}
\end{figure}

Another useful application of the \frameworkname{} is the use of HOCBFs with explicit ASIF RTA. While $\varphi_1(\boldsymbol{x})$, $\varphi_2(\boldsymbol{x})$, and $\varphi_4(\boldsymbol{x})$ all have a relative degree of 2, their conversion to control invariant constraints \cite{dunlap2021comparing,hibbard2022guaranteeing} caused them to become relative degree 1. To show the usefulness of HOCBFs, another scenario was simulated where $\varphi_1(\boldsymbol{x})$ was not converted to become control invariant, and was instead converted to become a HOCBF. While this does not guarantee control invariance, the constraint can still be enforced through an appropriate choice of $\alpha(x)$. The chief collision avoidance constraint for this simulation is shown in Figure \ref{fig:hocbf_collision}, where the explicit ASIF RTA assures safety of the constraint for the entire simulation. The average computation time across 100 simulations was 1.45 seconds per simulation, which is approximately the same as the previous simulation. Using HOCBFs allows the designer to compute non-trivial gradients without significant increases in computation time.

\section{CONCLUSION}
The \frameworkname{} presented in this paper is a powerful toolkit for building advanced RTA modules to provide continuous safety assurance to autonomous systems. It proliferates advancements from the RTA community by providing simple procedures for implementation with minimal user effort. Through its strategic utilization of JAX automatic differentiation, state-of-the-art Simplex and ASIF RTA techniques can be universally applied to many control problems by specifying constraints, system dynamics, and occasionally backup controllers. This will greatly improve the adoption of RTA techniques and accelerate the expansion of autonomous systems.

Case studies in multi-agent spacecraft inspection were presented demonstrating the efficacy and flexibility of the framework's constraint-based RTA implementations. Rather than manually deriving RTA implementations, cutting edge ASIF RTA algorithms were automatically applied to these problems. Constraints and dynamics were easily composed and recycled between approaches while no convoluted derivatives were hand-computed. With this scalable and extensible framework, RTA can be applied in the same manner to a multitude of problems within the autonomous systems space.





\bibliographystyle{Bibliography/IEEEtran}
\bibliography{Bibliography/IEEEabrv,Bibliography/root}

\end{document}